\documentclass[aps,prl,twocolumn,showpacs,groupedaddress]{revtex4}
\usepackage{amsmath}
\usepackage[latin1]{inputenc}
\usepackage[dvips]{graphicx}
\usepackage{amssymb}
\usepackage{color}
\newcommand{\T}{{\cal T}}
\newcommand{\R}{{\cal R}}
\newcommand{\V}{{\cal V}}

\begin{document}
%\preprint{}
\title{Quantum coherence engineering in the integer quantum Hall regime}
\author{P-A. Huynh$^1$,  F. Portier$^1$, H. le Sueur$^2$, G. Faini $^2$, U. Gennser $^2$, D. Mailly $^2$, F. Pierre $^2$, W. Wegscheider$^3$, and P. Roche$^1$}\email{patrice.roche@cea.fr}

\affiliation{$^1$CEA, SPEC, Nanoelectronics group, URA 2464, F-91191 Gif-sur-Yvette, France}
\affiliation{$^2$CNRS, LPN,Phynano Team, route de Nozay, 91460 Marcoussis, France}
\affiliation{$^3$Laboratory for Solid State Physics, ETH Z\"{u}rich, CH-8093 Z\"{u}rich, Switzerland}.
\date{\today}

\begin{abstract}
We present an experiment where the quantum coherence in the edge states of the integer quantum Hall regime is tuned with a decoupling gate. The coherence length is determined by measuring the visibility of quantum interferences in a Mach-Zehnder interferometer as a function of  temperature, in the quantum Hall regime at filling factor two. The temperature dependence of the coherence length can be varied by a factor of two. The strengthening of the phase coherence at finite temperature is shown to arise from a reduction of the coupling between co-propagating edge states. This opens the way for a strong improvement of the phase coherence of Quantum Hall systems. The decoupling gate also allows us to investigate how inter-edge state coupling influence the quantum interferences' dependence on the injection bias.  We find that the finite bias visibility can be decomposed into two contributions: a Gaussian envelop which is surprisingly insensitive to the coupling, and a beating component which, on the contrary, is strongly affected by the coupling.
\end{abstract}
\maketitle

Rare are the cases where quantum coherence can simply be controlled with a knob. This is because phase coherence is generally limited by the coupling of the system to its environment, and that this coupling is not easily controlled. In a two-dimensional electron gas in the integer quantum Hall (IQH) regime such a control is possible, due to the simplicity of the environment. In this regime, electrical transport occurs through one-dimensional chiral channels localized on the edges of the electron gas. Chirality reduces electron scattering, increasing the electron coherence length \cite{Martin90PRL64p1971}. This has motivated recent theoretical proposals to use these edge sates for quantum information experiments, and has renewed the interest of the community for precise investigations of quantum coherence and energy relaxation in the IQH regime. When two edge states are present, they constitute their own mutual environment. More specifically, the thermal charge noise in the one limits the phase coherence in the other \cite{Roulleau08PRL100n126802,Roulleau08PRL101n186803}. Taking advantage of this, we have designed a new Mach-Zehnder interferometer where we measure quantum interferences in the outer edge state while controlling the trajectory of the inner one with additional gates. 
This allows us to tune the coupling between edge states, resulting in an unprecedented way to control the coherence in the IQH regime. Our measurements show that one can increase the coherence length by nearly a factor two.
\begin{figure}
\includegraphics[angle=-90,width=9cm,keepaspectratio,clip]{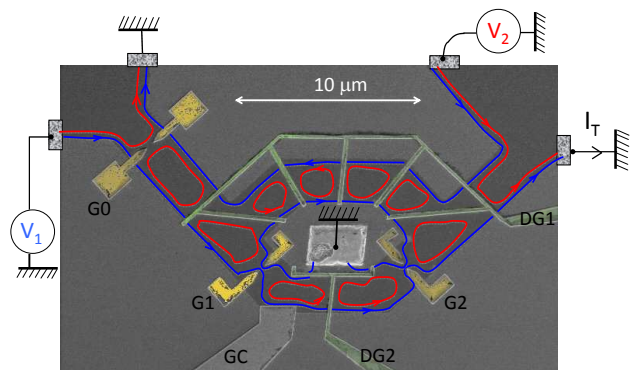}
\caption{(color online) Scanning Electronic Microscope view of the sample: the quantum point contacts (QPC) $G1$ and $G2$ are the beam splitters of the MZI.  Additional gates $DG1$ and $DG2$ are placed on the upper and lower arms
to force the inner edge state into small closed loops. The interferences are realized on the outer edge state.
The two edge states are fed with different bias voltage with the aid of $G0$: $V1$ for the
outer one and $V2$ for the inner one. The variation of the phase with respect to $V2$ allows us to determine the coupling between
the two edge states. The picture was taken before the final fabrication
step, where bridges are realized to connect both sides of the QPCs and the 
small ohmic contact in the center of the figure.}
\end{figure}

The Integer Quantum Hall regime  is obtained by applying a high magnetic field perpendicular to a two-dimensional gas
at low temperature. When the number of electrons per quantum of flux (the filling factor) is an integer, the electrical transport occurs through one-dimensional chiral modes on the edge of the sample: the edge states. The number of these edge states is equal to the filling factor. This one-dimensional  chiral transport has made new quantum experiments with electrons possible. For example, quantum interference experiments in the IQH regime have allowed the first observation of two electron interferences \cite{Neder07Nature448p333}, a first step toward the observation of the violations of Bell's inequalities \cite{Samuelsson04PRL92n026805}. Alternatively, combining a single electron gun \cite{Feve07Science316p1169} with a Mach-Zehnder interferometer \cite{Ji03Nature422p415} could permit the realization of Wheeler's delayed choice Gedanken experiment \cite{Jacques07Science315p966} with electrons. However, these electronic versions of optical experiments suffer from one major problem: an electron carries a charge with which it interacts with the surrounding world, leading to a finite quantum coherence length and
a finite energy exchange length. Recently, both lengths have been measured in the IQH regime at filling factor two, definitively showing the role of the interaction between the inner and outer edge state
 \cite{Lesueur10PRL105n056803,Roulleau08PRL100n126802,Roulleau08PRL101n186803}. While the dependence of the coherence length with temperature has been clearly identified to result from the thermal charge noise in the neighboring edge state
 \cite{Roulleau08PRL101n186803}, it has not yet been possible to clearly identify the role of energy exchanges on the repeatedly observed but poorly understood Gaussian shape of the visibility as a function of the bias voltage  \cite{Roulleau07PRB76n161309, Litvin08PRB78n075303,Yamauchi09PRB79n161306}. It has been demonstrated recently that the energy exchange between the edge states, which form at filling factor two, can be frozen by opening a gap in the excitations of the inner edge state (IES). This is done by forcing the IES on small closed loop trajectory of length $L_\delta$ of the order of 8~$\mu$m, leading to an energy spacing $E_\delta\sim hv_D/L_\delta\sim50$~$\mu$eV, and hence freezing energy exchange below this value \cite{Altimiras10PRL105n226804}. Inspired by theses findings, we have designed a new MZI with additional gates  $DG1$ and $DG2$ used to localize the IES on loops typically 8~$\mu$m long (see Fig.1). The goal of these new gates is twofold: first, to freeze the charge fluctuations in the nearby environment and hence increase the finite temperature coherence length $l_\varphi$; second, to check if finite bias energy exchanges are involved in the finite bias visibility decrease. One would expect in this case an enhancement of the robustness of quantum interference with the bias. While we do clearly observe an enhancement of the finite temperature coherence length that we prove to be due to a reduction of the coupling to the environment, the robustness of the visibility with the bias voltage is surprisingly poorly affected by the decoupling gates.
 
The MZI is realized with a high mobility two dimensional electron gas with a density of 1.1~10$^{11}$ cm$^{-2}$ and a mobility of 3~10$^6$ cm$^2$V$^{-1}$s$^{-1}$. The measurements have been performed at filling factor two with a 2.63~T magnetic field. The two beam splitters of the MZI are Quantum Point Contacts $G1$ and $G2$ with transmission probability of the outer edge state (OES) $\T_1$ and $\T_2$ (the IES is fully reflected). Each arm of the MZI is $11.8$~$\mu$m long. The differential transmission $\T=dI_T/dI_0$, $I_T$ being the transmitted current and $I_0$ the incoming current, is measured with standard lock-in techniques with a 2 $\mu$V$_{RMS}$ excitation. Thanks to the additional gate $G_0$, the IES and OES can be fed with different biases $V_1$ and $V_2$ respectively. The interference pattern is revealed by ramping the voltage on the side gate $G_C$, which changes the Aharonov-Bohm
flux $\phi$ through the area defined by the two arms of the interferometer
$\T=\T_{mean}(1+\V\sin(\varphi))$,
$\V$ being the visibility proportional to $\exp(-T/T_\varphi)\sqrt{\T_1\R_2\T_2\R_1}/(\T_1\T_2+\R_1\R_2)$. $T$ is the temperature, 
$\R_i=1-\T_i$, and $\varphi=2\pi\phi/\phi_0$ where $\phi$ is the magnetic flux through 
the area defined by the two arms of the MZI.
%It has been verified that the cross talk between gates $DG_1$, $DG_2$, $G_1$, and $G_2$ only play a minor role on the gate tuning.
Both gates $DG_i$ fully transmit the IES at 0.3~V and fully reflect it at 0.1~V.
 
%the non symmetric behavior of the visibility with respect to the bias can find result from a non symmetric
%behavior of the transmission probabilities .... we could check this ?   
\begin{figure}
\includegraphics[angle=-90,width=8.5cm,keepaspectratio,clip]{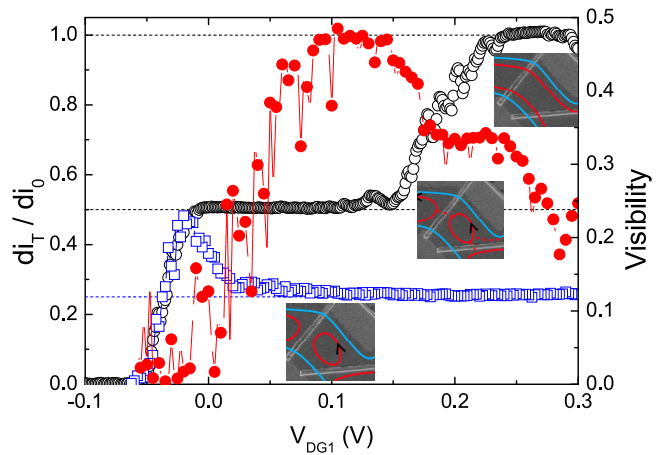}
%\includegraphics[width=8.5cm,keepaspectratio,clip]{Figure2Newb.ps}
%poster5.opj
%An alternative for the plot is the one extracted from PhDopj2corr.opj in ANALYSE/11_03poster (figure2B.ps)
%Another alternative plot realised with the side gate to reveal interferences is in Analyse/11_03poster/PhDopj1 (the last choice)
\caption{(color online) Solid (red) dot: Visibility of quantum interferences for different values of $V_{DG_1}$ as a function of the 
bias voltage at 25 ~mK. Open (black) dot: Transmission probability the upper arm of the interferometer. Open (blue) square: mean transmission through the MZI when revealing interference, the IES being fully reflected by G0. The departure from the 0.25 value below $V_{DG_1}\sim0.1$~V indicates a detuning of the
MZI due to $DG_1$. The trajectories of both edge states  are shown schematically for different values of $V_{DG_1}$ in
the inserted SEM pictures.}
\end{figure}

Figure 2 shows the visibility $\V$ of quantum interferences at the base temperature of 25~mK, the transmission probability through the upper arm, and the transmission through the MZI averaged over ten Aharonov-Bohm periods.  One can see that $DG_1$ has a great impact on the visibility, which is enhanced by a factor of the order of two between full transmission and full reflection of the OES. Acting on $DG_2$ also increases the visibility, but much less.
%One can notice on figure 2 that  there is no sudden rising of the visibility coinciding with the full reflection of the IES: 
The visibility saturates when the IES is fully reflected, before decreasing at lower $V_{DG_1}$. This decrease is most probably due to the deformation of the OES leading to an imbalance in the two arms trajectory length and/or the detuning of the MZI, which is illustrated by the departure of the mean transmission from the 0.25 value (squares on figure 2). 

\begin{figure}
\includegraphics[angle=-90,width=8.5cm,keepaspectratio,clip]{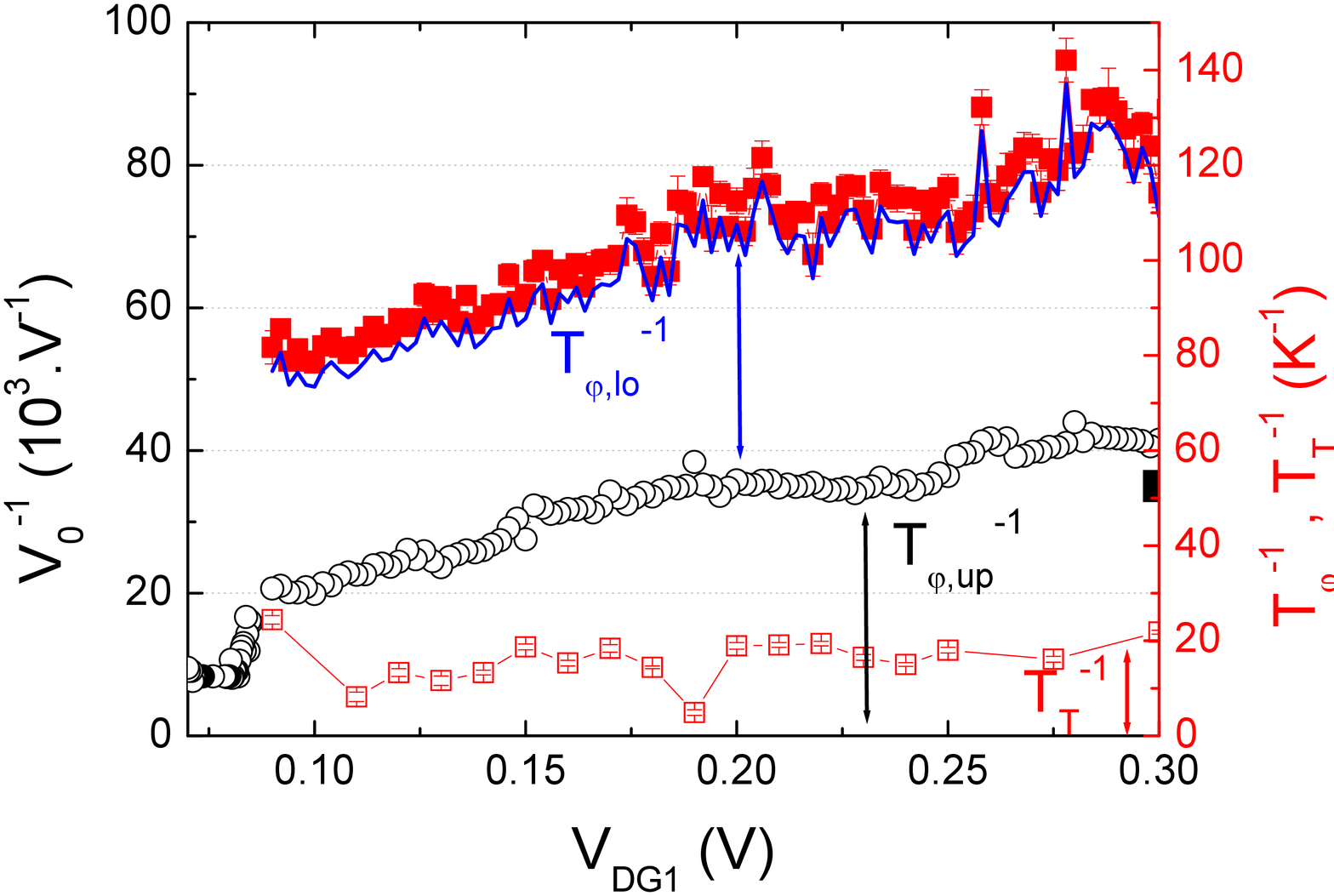}
\includegraphics[angle=-90,width=8cm,keepaspectratio,clip]{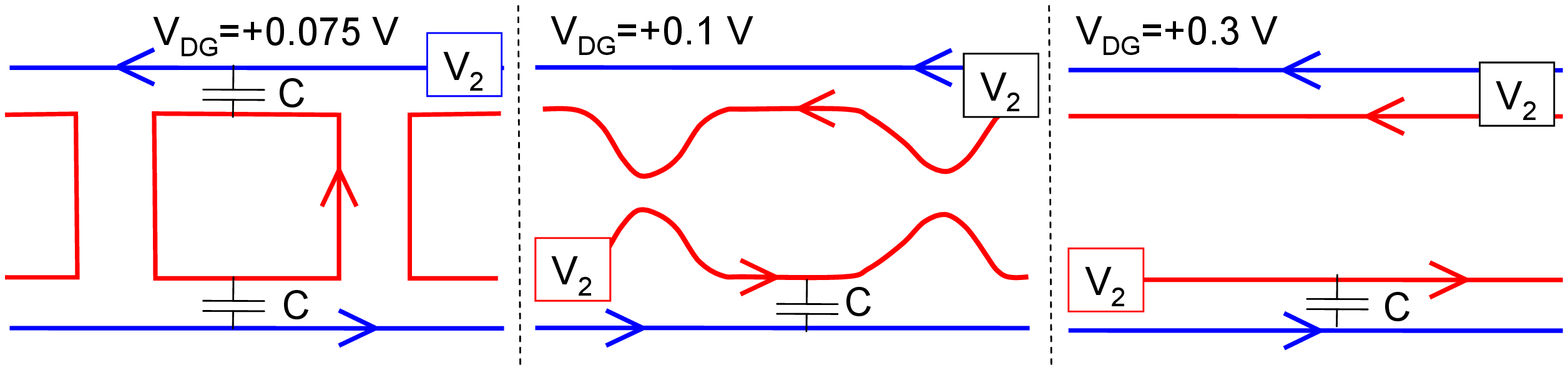}
%poster2.opj .. figure4c.ps
\caption{
%(color online) Left axe: open circles: Coupling $V_0^{-1}$ between the IES and the OES. Right axe: filled square: Measured decoherence $T_{\varphi}^{-1}=-\partial\ln(\V)/\partial T$ as a function of V$_{DG_1}$. Open squares : calculated $T_T^{-1}$ from the phase variation of the bias (see text). Solid black square: expected dephasing in the upper arm deduced from a which path experiment (see text). Solid line: calculated dephasing $T_{\varphi}^{-1}$ after subtracting the thermal averaging \cite{EPAPS}. The lower panel is a sketch of the edge state coupling for different values of $V_{DG}$.}
(color online)  Open circles: Coupling $V_0^{-1}$ between the IES and the OES (left axis); filled squares (right axis): measured dephasing $T_{\varphi}^{-1}=-\partial\ln(\V)/\partial T$; solid line (right axis): measured $T_{\varphi}^{-1}$ after subtracting the thermal averaging \cite{EPAPS}; open squares (right axis):  $T_T^{-1}$, calculated from the phase variation of the bias (see text).  The solid black square at $V_{DG1}=0.3~V$ represents the dephasing $T_{\varphi,up}^{-1}$ in the upper arm, deduced from a which path experiment.  
The scale of the left axis has been adjusted so that the dephasing $T_{\varphi,up}^{-1}$ of the upper arm corresponding to  $V_0^{-1}$ can be read on the right axis.
The lower panel is a sketch of the edge state coupling for different values of $V_{DG1}$.} 
\label{Dec}
\end{figure}

We wish to stress that the effect of $DG_1$ is not a reduction of the thermal smearing which could occur in case of an imbalance of the time of flight through the two arms. Thermal smearing leads to a visibility dependence $\V\propto T\sinh^{-1}(T/T_T)\sim \exp(-T/T_T)$ for $T\gg T_T$ with $T_T^{-1}=\pi k_B/e\times \partial\varphi / \partial V_{DS}$, where  $\partial\varphi / \partial V_{DS}$ is phase dependance of the Aharonov-Bohm oscillations as a function of the dc bias applied on the MZI \cite {Chung05PRB72p125320,Roulleau08PRL101n186803}. $T_T^{-1}$ determined from the variation of the  AB phase with the bias is represented in figure 3. $T_T^{-1}$, which is of the order of  20~K$^{-1}$, has a negligible impact on the visibility $\V$. Figure 3 illustrates the underlying physics leading to the coherence enhancement. It proves
that the coherence is modified because $DG_1$ allows us to change the coupling of the OES with its environment. Figure 3 is constructed in the following way. We first measure the temperature dependence of $\V$ for different values of $V_{DG_1}$. It shows an exponential behavior $\V\propto\exp(-T/T_\varphi)$, with the decoherence $T_\varphi^{-1}$ decreasing when decreasing  $V_{DG_1}$.
The inter-edge state coupling, $V_0^{-1}=(2\pi^{-1})\partial\varphi/\partial V_2$ is simultaneously measured, using the 
method of ref.\cite{Roulleau08PRL101n186803}: one feeds the IES with a potential $V_2$ while the OES remains at equilibrium. The IES plays the role of a side gate used to reveal interferences in the OES.

The variations of $T_\varphi^{-1}$ and $V_0^{-1}$ with $V_{DG_1}$ are remarkably similar, strongly suggesting that these two quantities have a common microscopic origin. Assuming that the dephasing is a sum of the dephasings in the upper and the lower arms, $T_\varphi^{-1}= T_{\varphi,up}^{-1} + T_{\varphi,lo}^{-1}$, 
with  $T_{\varphi,up}^{-1} = \alpha  V_0^{-1}$ following \cite{Roulleau08PRL101n186803}, one can set the proportionality factor $ \alpha$ so that the variations of $V_0^{-1}$ reproduce those of $T_\varphi^{-1}$. The scales of Fig. 3 are adjusted following
this procedure, such that the open black circles represent $V_0^{-1}$ to the left and the corresponding $T_{\varphi,up}^{-1}$ to the right. Independently, we deduced $T_{\varphi,up}^{-1}$ from a 'which path' experiment, where the AB interferences of the outer edge state were washed out by shot noise produced in the inner edge state. This was done for $V_{DG1}=$~0.3~V by setting the transmission of the IES on $G0$ at one half and  applying various bias voltages on $V_2$. Following \cite{Roulleau08PRL101n186803}, we can extract $T_{\varphi,up}^{-1}$  from the observed variations of  the visibility with $V_2$. The obtained value, $51\pm2$~K$^{-1}$, represented by the solid black square in Fig. 3, is in remarkable agreement with the estimation based on the variations of $V_0^{-1}$. Note that this experiment is \emph{not} possible for lower values of  $V_{DG_1}$, when the decoupling gate DG depletes the 2DEG, as this introduces partial reflections of the inner edge state. 
%In figure 3 we have set the left and right scale 
%such that $\delta T_{\varphi}^{-1}\propto \delta V_0^{-1}$. As a result, the evolution of the
%dephasing  in the upper arm $T_{\varphi,up}$ corresponds to the black circles on the right scale. This procedure leads to a $T_{\varphi,up}$ which coincide remarkably with the deduce upper arm decoherence (black square at $V_{DG_1}=0.3$~V). This later decoherence being calculated after realizing a which-path experiment following the method of  ref.\cite{Roulleau08PRL101n186803,EPAPS}. 
\begin{figure}
\includegraphics[angle=-90,width=7cm,keepaspectratio,clip]{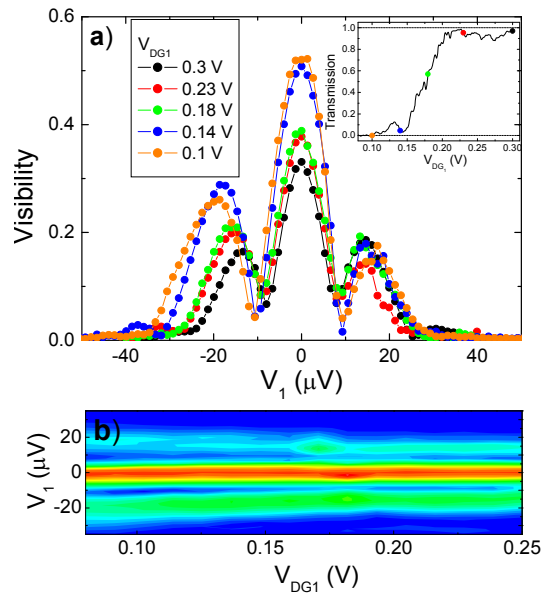}
%poster5.opj et la partie 2D extraite de PhDopj6simpl.opj
\caption{(color online) a) Visibility as a function of the bias voltage for different polarization of $DG1$. Inset: transmission of the IES through the upper arm of the MZI as a function of $V_{DG1}$. b) 2D color plot of the visibility normalized to the zero bias visibility as a function of $V_{DG1}$ and $V_1$.}
\label{Dec}
\end{figure}
As expected, the difference between $T_\varphi^{-1}$ and the adjusted $T_{\varphi,up}^{-1}$, corresponding to the dephasing in the lower arm,  is almost independent of DG1.

We now focus on the behavior at full reflection of the IES (around $V_{DG1}=0.08$~V). 
The coupling $V_0^{-1}$ changes abruptly by a factor of the order of two. 
As sketched in Fig. 3, lower panel, the decrease of the coupling
is not due to a variation of the actual coupling between the two edge states. In fact, here 
the IES is no longer at the potential $V_2$. The measured variation
of the AB phase with $V_2$ results from the coupling with the counter-propagating
OES at potential $V_2$ through the small loop formed by the IES. This
process leads to a factor two in the coupling, corresponding to
two geometrical capacitances in series, each mimicking the local coupling between the two neighboring edge states.

Unexpectedly, the base temperature visibility reaches a maximum (around $V_{DG1}=0.1$~V) before decreasing (see fig.2), 
indicating that the variations of $V_0$ no longer imply a variation of $T_\varphi$ \footnote{While we have
not measured the temperature dependence of $\V$ when the loops are formed,
for higher gate voltages the base temperature visibility has systematically varied like $T_\varphi^{-1}$.}.
Following the approach of ref. \cite{Roulleau08PRL101n186803,Seelig01PRB64n245313}, 
only charge fluctuations on a time scale longer than the time
of flight through the MZI account for the dephasing. Here, the static charge in the small loops should be frozen and electron-hole
excitations are expected to occur only at temperatures larger than $E_\delta/k_B$, eliminating low temperature charge fluctuations 
on the small loops. Hence, $T_\varphi^{-1}$ should result from thermal charge noise in the counter
propagating OES coupled to the interfering OES
through the loops: $T_\varphi^{-1}$ should be proportional to $V_0^{-1}$, even when the loops are formed. 
The contradiction between this simple model and our observations may result from different causes.
For example, the freezing of charge fluctuations could be compensated by an imbalance of electron trajectories
in both arms leading to thermal averaging; or the drift velocity may be overestimated, leading to an overestimation of the energy gap of the loops. Also, it could be that the charge noise due to the dissipative part of the coupling between the interfering edge state and the metallic gates leads to decoherence \cite{Seelig01PRB64n245313}, or that another mechanism implying coupling between edge states \cite{Levkivskyi08PRB78n045322} is responsible for decoherence.  For the time being, we lack experimental data to answer to these questions.
%%%%%%%%%
Another aspect that we address in this experiment is the role of inter edge state coupling and energy exchange 
\cite{Lesueur10PRL105n056803} on the finite bias visibility. Although several theoretical works  have attempted to 
explain the unexpected multiple lobe structure observed in the variations of the visibility with bias voltage \cite{Neder06PRL96p016804,Roulleau07PRB76n161309,Litvin08PRB78n075303}, so far no scenario has been fully validated experimentally. In particular, the Gaussian envelop of the visibility $\V\propto \exp(-V^2/2V_l^2)$ revealed in ref.\cite{Roulleau07PRB76n161309} and also observed in Fabry-P{\'e}rot interferometers \cite{Yamauchi09PRB79n161306} has not been accounted for. 
\begin{figure}
\includegraphics[angle=-90,width=7cm,keepaspectratio,clip]{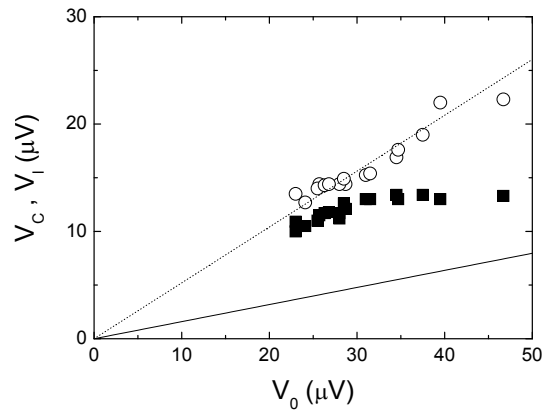}
%PhD9Last.opj
\caption{$V_C$ (open dots) and $V_l$ (black squares) as a function of the coupling between the two
edge states characterized by $V_0$. The dashed line is a guide for the eyes. The solid line is the theoretical prediction \cite{Levkivskyi08PRB78n045322} assuming the same inter edge state coupling in both arms of the interferometer.} 
\end{figure}
In figure 4, the visibility as a function of the bias is plotted for different values of $V_{DG1}$. As already observed \cite{Neder06PRL96p016804,Litvin08PRB78n075303}, a side lobe structure shows up, with a first lobe around $V_1=20$ $\mu$V, and a much smaller one around $V_1 \simeq 40$~$\mu$V. When the zero bias visibility is enhanced
by decreasing $V_{DG1}$, the width of the central lobe $\sim V_l$ is hardly affected, even when the IES is fully reflected ($V_{DG1}\sim0.08$~V. This result points towards the fact that if indeed \cite{Altimiras10PRL105n226804}
energy exchanges are frozen, they are not the main process leading to the finite bias visibility decrease. It also shows that 
the inter edge coupling is not involved in the mechanism leading to
the Gaussian envelop.
An additional information derived from this measurement is the evolution of 
the second minimum of the visibility \cite{EPAPS}. 
In the present experiment, as we are able to control the coupling between 
the edge states, we can make a comparison between our measurements and the theory of Levkivskyi and
Sukhorukov \cite{Levkivskyi08PRB78n045322}
where the coupling between edge states is the basic ingredient to explain the presence of the multiple side lobe structure.
In this theory, a charge excitation in one edge state is decomposed into two coupled modes, a neutral one (with speed $v$)
and a charged one (with speed $u$) delocalized on the two edge states. One expects $u\gg v$. There is a simple relation
between our coupling parameter
$V_0$ and $v$: $eV_0=\pi \hbar v/L$, assuming the upper and lower arms of the interferometer have equal lengths $L$. 
Beating between the two modes leads to a visibility of quantum interferences $\propto |\cos(eV_1 L/(2\hbar v))|=|\cos(V_1/V_C)|$
with $V_C=V_0/(2\pi).$ The finite bias visibility that we measure is very well fitted by a combination
of a Gaussian envelop times a cosine term
 \cite{Litvin08PRB78n075303,EPAPS}. The first minimum is mainly determined by $V_l$, poorly affected by the coupling between edge sates (see Figure 5). The second minimum is mainly determined by the cosine term. In figure 5 one can notice that $V_C$ is proportional to the coupling $V_0$ with a proportionality factor of the same order of magnitude as the predictions of \cite{Levkivskyi08PRB78n045322}. It is however difficult to make a quantitative comparison as we control the coupling in only one arm of the MZI. Nonetheless, these results
definitively show that the underlying mechanism leading to the higher order minima of the side lobe structure 
involves the interaction between the two edge states.
%PhDopj9.opj

To summarize, we have strongly enhanced the quantum coherence in the integer quantum Hall regime at filling
factor two by protecting the interfering edge state from the thermal charge noise of its environment. Two components of the finite bias visibility has been identified: a Gaussian envelop, poorly affected by the inter edge state coupling 
and, a beating term strongly dependent on the coupling between edge state as recently proposed by a theory. 

PR would like to thank E. Sukhorukov for simulating discussions. This work has been supported by
the ANR grant IQHAR (ANR-2011-BS04-022-01).

%\bibliography{References-1}

\begin{thebibliography}{10}

\bibitem{Martin90PRL64p1971}
T. Martin and S. Feng, Phys. Rev. Lett. {\bf 64},  1971  (1990).

\bibitem{Roulleau08PRL100n126802}
P. Roulleau {\it et~al.}, Phys. Rev. Lett. {\bf 100},  126802  (2008).

\bibitem{Roulleau08PRL101n186803}
P. Roulleau {\it et~al.}, Phys. Rev. Lett. {\bf 101},  186803  (2008).

\bibitem{Neder07Nature448p333}
I. Neder {\it et~al.}, Nature {\bf 448},  333  (2007).

\bibitem{Samuelsson04PRL92n026805}
P. Samuelsson, E.~V. Sukhorukov, and M. B{\"u}ttiker, Phys. Rev. Lett. {\bf
  92},  026805  (2004).

\bibitem{Feve07Science316p1169}
G. F\'{e}ve {\it et~al.}, Science {\bf 316},  1169  (2007).

\bibitem{Ji03Nature422p415}
Y. Ji {\it et~al.}, Nature {\bf 422},  415  (2003).

\bibitem{Jacques07Science315p966}
V. Jacques {\it et~al.}, Science {\bf 315},  966  (2007).

\bibitem{Lesueur10PRL105n056803}
H. Lesueur {\it et~al.}, Phys. Rev. Lett {\bf 105},  056803  (2010).

\bibitem{Roulleau07PRB76n161309}
P. Roulleau {\it et~al.}, Phys. Rev. B {\bf 76},  161309(R)  (2007).

\bibitem{Litvin08PRB78n075303}
L.~V. Litvin {\it et~al.}, Phys. Rev. B {\bf 78},  075303  (2008).

\bibitem{Yamauchi09PRB79n161306}
Y. Yamauchi {\it et~al.}, Phys. Rev. B {\bf 79},  161306(R)  (2009).

\bibitem{Altimiras10PRL105n226804}
C. Altimiras {\it et~al.}, Phys. Rev. Lett {\bf 105},  226805  (2010).

\bibitem{EPAPS}
See supplementary material.

\bibitem{Chung05PRB72p125320}
V.~S.-W. Chung, P. Samuelsson, and M. B{\"u}ttiker, Phys. Rev. B {\bf 72},
  125320  (2005).

\bibitem{Seelig01PRB64n245313}
G. Seelig and M. Buttiker, Phys. Rev. B {\bf 64},  245313  (2001).

\bibitem{Levkivskyi08PRB78n045322}
L.~P. Levkivskyi and E.~V. Sukhorukov, Phys. Rev. B {\bf 78},  045322  (2008).

\bibitem{Neder06PRL96p016804}
I. Neder {\it et~al.}, Phys. Rev. Lett. {\bf 96},  016804  (2006).

\end{thebibliography}

\end{document}